\documentclass{eptcs}
\usepackage{breakurl}
\usepackage{graphicx}
\usepackage{paralist}

\usepackage{color}
\definecolor{lstbg}{rgb}{0.98,0.98,0.98}

\usepackage{listings}
\lstdefinelanguage{gretl}{
  morekeywords={E, V, as, bag, eSubgraph, end, exists!, exists, false, forall,
    from, import, in, let, list, null, path, pathSystem, rec, report,
    reportBag, reportMap, reportSet, thisEdge, thisVertex, map,
    role,    
    set, store,
    true, tup, using, vSubgraph, where, with, to},
  morekeywords={new, protected, public, static, final, private, if, switch,
    case, for, break, default, throw, void, else, return, continue},
  morekeywords={[2]and, avg, contains, containsKey, count, degree, depth,
    difference, distance, dividedBy, edgesConnected, edgesFrom, edgesTo,
    edgeTrace, edgeTypeSet, endVertex, enumConstant, equals, extractPath,
    getEdge, get, getValue, getVertex, grEqual, grThan, hasAttribute, hasType,
    id, inDegree, innerNodes, intersection, isAcyclic, isEmpty, isA, isCycle,
    isIn, isIsolated, isLoop, isNeighbour, isNull, isParallel, isReachable,
    isSibling, isSubPathOfPath, isSubSet, isSuperSet, isTrail, isTree, keySet,
    leaves, leEqual, leThan, matches, maxPathLength, minPathLength, minus,
    modulo, nequals, nodeTrace, not, nthElement, theElement, or, outDegree,
    package-info, parent, pathConcat, pathLength, pathSystem, plus, pos,
    reachableVertices, recordInstance, reMatch, schemaFunctions, siblings,
    squareRoot, startVertex, subtypes, sum, supertypes, symDifference, times,
    toString, type, typeName, typeSet, types, uminus, union, values,
    vertexTypeSet, weight, xor, error},
  emph={AddMappings, Assert, CreateSubgraph, CreateVertexClass,
    CreateAbstractVertexClass, CreateEdgeClass, CreateAbstractEdgeClass,
    AddSubClass, AddSubClasses, AddSuperClass, AddSuperClasses, CopyDomain,
    CreateAttribute, CreateAttributes, CreateListDomain, CreateSetDomain,
    CreateMapDomain, CreateRecordDomain, CreateEnumDomain, CreateVertices,
    CreateEdges, ExecuteTransformation, MatchReplace, MergeVertices, Delete,
    SetAttributes, RedefineFromRole, RedefineFromRoles, RedefineToRole,
    RedefineToRoles, transformation, Iteratively},
  emphstyle=\bf\underbar,
  sensitive=true,
  string=[b]{\\"},
  morecomment=[l]{//}}

\makeatletter
\lst@AddToHook{TextStyle}{\let\lst@basicstyle\footnotesize\sffamily}
\makeatother

\newcommand{\myemail}[0]{\email{horn@uni-koblenz.de}}
\lstdefinestyle{constfold}{language=gretl, name={constfold}, firstnumber=auto}
\lstdefinestyle{is}{language=gretl, name={instrselect}, firstnumber=auto}

\def\titlerunning{Solving the TTC 2011 Compiler Optimization Case with GReTL}
\def\authorrunning{Tassilo Horn}

\title{Solving the TTC 2011 Compiler Optimization Case with GReTL}
\author{Dipl.-Inform. Tassilo Horn\\
  \myemail\\
  Institute for Software Technology\\
  University Koblenz-Landau, Campus Koblenz}

\begin{document}
\maketitle
\begin{abstract}
  This paper discusses the GReTL solution of the TTC 2011 \emph{Compiler
    Optimization} case \cite{compileroptimizationcase}.  The submitted solution
  covers both the \emph{constant folding} task and the \emph{instruction
    selection} task.  The \emph{verifier} for checking the validity of the
  graph is also implemented, and some additional test graphs are provided as
  requested by the \emph{extension}.
\end{abstract}

\section{Introduction}
\label{sec:introduction}

GReTL (Graph Repository Transformation Language, \cite{gretl-icmt2011}) is the
operational transformation language of the TGraph technological space
\cite{tgapproach08}.  Models are represented as typed, directed, ordered, and
attributed graphs.  GReTL uses the GReQL (Graph Repository Query Language,
\cite{FestschriftNagl2010}) for its matching part.

The standard GReTL operations are currently targeted at out-place
transformation which create the target metamodel and the target graph
simultaneously, but this case requires changing the graph in-place.  Luckily,
GReTL is designed as an extensible language.  Adding custom operations requires
only specializing one framework class, overriding one method that implements
the operation's behavior, and implementing another factory method that is
responsible for creating an instance of the new operation initialized with the
given parameters.  Consequently, some in-place operations have been added to
the language.  There is one operation \emph{MatchReplace} which is similar to a
rule in graph replacement systems.  There is an operation \emph{Delete} for
deleting elements, an operation \emph{MergeVertices} for merging a set of
duplicate vertices into one canonical representative, and a higher-order
operation \emph{Iteratively} which receives another operation and applies it as
long as possible.  The implementation of the most complex one,
\emph{MatchReplace}, is about 380 lines of Java code including comments and
empty lines, the other three operations are at most 80 lines or less.  Thus,
this case was solved using a mixed bottom-up and top-down approach: the
language was extended with three new operations, and then these operations were
used to write the transformations.

\section{Case Solutions}
\label{sec:task-solutions}

In this section, the two transformation tasks are discussed in sequence.  The
solutions can be run on the SHARE image \cite{share-demo}.

\subsection{Constant Folding}
\label{sec:constant-folding}

The GReTL solution for the constant folding task does a bit more than what was
required.  It contains operation calls that realize the following
optimizations:

\begin{compactenum}[(1)]
\item \textsf{Binary} operations depending only on two \textsf{Const} vertices
  are replaced with with a \textsf{Const} containing the result of evaluating
  the binary.
\item \textsf{Not} operations depending only on a \textsf{Const} are similarly
  replaced by the evaluation result.
\item Unused \textsf{Const} vertices are deleted.
\item \textsf{Const} vertices representing the same value are merged into one.
\item \textsf{Cond} nodes (conditional jumps) depending on a \textsf{Const} are
  replaced with unconditional jumps (\textsf{Jmp}).
\item Unreachable code is eliminated.
\item The operand edges of \textsf{Phi} vertices are renumbered if needed.
\item \textsf{Phi} vertices with only one argument are replaced by a direct
  dependency between the argument and the vertex depending on that
  \textsf{Phi}.
\item \textsf{Block} vertices containing only an unconditional \textsf{Jmp} are
  removed.
\item In subgraphs consisting of commutative and associative binary operations
  (\textsf{Add} or \textsf{Mul}) constants are pulled up in order to generate
  new possibilities for the constant folding optimization (1), e.g., a graph
  structure $((1 \star x) \star 2)$ is replaced with $((1 \star 2) \star x)$
  for $\star \in \{+, *\}$.
\end{compactenum}

In the following, the optimizations (1), (3), and (4) are discussed, starting
with the evaluation of binary operations depending only on constants.  A helper
function is defined that receives some \textsf{Binary} operation \textsf{bin}
and the values of its left and right operand (\textsf{lval}, \textsf{rval}) and
returns the evaluation result.

\begin{lstlisting}[style=constfold]
evaluateBinary() := using bin, lval, rval:
    hasType{Add}(bin)  ? lval + rval
  : hasType{Sub}(bin)  ? lval - rval
  : hasType{Mul}(bin)  ? lval * rval
  : hasType{Div}(bin)  ? (let d := lval / rval in d < 0 ? ceil(d) : floor(d))
  : hasType{Mod}(bin)  ? lval % rval
  : hasType{Shl}(bin)  ? bitOp("SHIFT_LEFT", lval, rval)
  : hasType{Shr}(bin)  ? bitOp("UNSIGNED_SHIFT_RIGHT", lval, rval)
  : hasType{Shrs}(bin) ? bitOp("SHIFT_RIGHT", lval, rval)
  : hasType{And}(bin)  ? bitOp("AND", lval, rval)
  : hasType{Or}(bin)   ? bitOp("OR", lval, rval)
  : hasType{Eor}(bin)  ? bitOp("XOR", lval, rval)
  : hasType{Cmp}(bin)  ?
        (bin.relation = "GREATER"        ? (lval > rval  ? 1 : 0)
       : bin.relation = "GREATER_EQUALS" ? (lval >= rval ? 1 : 0)
       : bin.relation = "LESS"           ? (lval < rval  ? 1 : 0)
       : bin.relation = "EQUAL"          ? (lval = rval  ? 1 : 0)
       : bin.relation = "NOT_EQUAL"      ? (lval <> rval ? 1 : 0)
       : bin.relation = "LESS_EQUAL"     ? (lval <= rval ? 1 : 0)
       : bin.relation = "TRUE"           ? 1
       : bin.relation = "FALSE"          ? 0
       : error("Don't know how to handle " ++ bin.relation))
  : error("Don't know how to handle " ++ bin);
\end{lstlisting}

Depending on the type of the given \textsf{Binary} \textsf{bin}, the correct
operation is dispatched using a sequence of conditional
expressions\footnote{GReQL conditional expression: \texttt{<testExp> ?
    <trueExp> : <falseExp>}} and applied to the operand values.  The division
has to be handled specially: in GReQL, a division results in a double value,
but here an integer division is intended.  To match these semantics, either the
floor or the ceiling of the result \textsf{d} is taken.

The next listing shows the replacement of all \textsf{Binary} vertices that
have two \textsf{Const} arguments with a new \textsf{Const} with \textsf{value}
set to the result of evaluating the binary operation.  The
\textsf{MatchReplace} operation has semantics similar but not identical to
rules in graph replacement systems.  Its first parameter is a template graph
that describes the structure of the subgraph to be created or changed (similar
to the RHS in rules in graph replacement systems), and following the arrow
symbol, there is a GReQL query that reports a set of matches.  The query is
comparable to the LHS in graph transformations, except that it calculates all
matches instead of one match at a time.  For all matches, the template graph is
applied, thereby skipping matches containing previously modified elements.

In the template graph, the dollar (\$) is a variable that holds the current
match.  Vertices are represented using paretheses and edges with arrows and
curly braces:
\begin{verbatim}
         (Type 'greql' | attr1 = 'greql2', attr2 = 'greql3')
      -->{Type 'greql' | attr1 = 'greql2', attr2 = 'greql3'}
\end{verbatim}

The semantics for vertices, but likewise for edges, are as follows.  The
optional \textsf{Type} is a vertex type name.  If no \textsf{Type} is specified
and \textsf{greql} evaluates to a vertex in the current match or any other
vertex in the graph, it is bound to the template vertex and preserved.  If
\textsf{Type} is given and \textsf{greql} evaluates to a vertex in the current
match, then that vertex is replaced with a new vertex of the specified type,
all edges incident to the replaced vertex are relinked to the new one, and the
new vertex is bound to the template vertex.  If \textsf{Type} is given but no
\textsf{greql} query, then a new vertex of the given type is created and bound
to that template vertex.  The attributes of the bound elements are set to the
result of the queries in the attribute list.  All vertices and edges in the
match that are not bound to any template graph element are deleted.  The
deletion of some vertex implies deletion of all incident edges.

\begin{lstlisting}[style=constfold]
  Iteratively MatchReplace ('startBlock') <--{Dataflow | position = '-1'}
                           (Const '$.op' | value = '$.value')
    <== from binOp: V{Binary}
        with hasType{Const}(lconst) and hasType{Const}(rconst)
        reportSet rec(op: binOp,
                      value: evaluateBinary(binOp, lconst.value, rconst.value),
                      edges: edgesFrom(binOp)) end
        where argEdges := sortByPosition(nonContainmentDfs(binOp)),
              lconst := endVertex(argEdges[0]),
              rconst := endVertex(argEdges[1]);;
\end{lstlisting}

The higher-order operation \textsf{Iteratively} applies the
\textsf{MatchReplace} operation as long as some match could be replaced.  The
query given to \textsf{MatchReplace} reports a set of matches where each match
is a record of a binary operation (\textsf{op}), the evaluation result gathered
by the \textsf{evaluateBinary()} helper (\textsf{value}), and the set of edges
starting at the binary operation (\textsf{edges}).

The template graph specifies that a \textsf{Const} has to be created with
\textsf{value} set to \textsf{\$.value}.  The new \textsf{Const} should be
connected to the \textsf{StartBlock}.  The connecting edge has to be of type
\textsf{Dataflow} with \textsf{position = -1}.  Because the new \textsf{Const}
refers to the binary operation \textsf{op} and additionally specifies a type,
it means the replacement of that binary with the new constant.  All edges
incident to the binary will be relinked to the constant.  Finally, the binary
\textsf{op} and all \textsf{edges} starting at it will be deleted, because they
occur in the match but not in the template graph.

This operation may have deleted vertices that were the single element depending
on some \textsf{Const}.  Thus, the following operation deletes all unused
\textsf{Const} vertices.

\begin{lstlisting}[style=constfold]
  Delete <== from c: V{Const} with inDegree(c) = 0 reportSet c end;
\end{lstlisting}

Furthermore, the operation may have created duplicate \textsf{Const} vertices,
i.e., constants with the same \textsf{value}.  The next operations merges them
in order to have exactly one \textsf{Const} per needed \textsf{value}.

\begin{lstlisting}[style=constfold]
  MergeVertices <== from const: V{Const}
                    reportMap const -> from d: V{Const}
                                       with d.value = const.value and d <> const
                                       reportSet d end
                    end;
\end{lstlisting}

The \textsf{MergeVertices} operation receives a query that has to result in a
map assigning to vertices a set of duplicates that should be merged.  Any
\textsf{Const} vertex is mapped to the set of all different \textsf{Const}
vertices with equal \textsf{value}.  Thus, for any used value there will be
exactly one canonical \textsf{Const}.

\subsection{Instruction Selection}
\label{sec:instruction-selection}

In this section, all operation calls for the instruction selection task are
discussed.  The first one transforms all \textsf{Binary} vertices with at least
one \textsf{Const} argument to immediate target binary operations.

\begin{lstlisting}[style=is]
MatchReplace (#'"Target" ++ typeName($.binary) ++ "I"' '$.binary' | value = '$.value',...)
 <== from b: V{Binary}
     with count(constEdges) > 0
     reportSet rec(binary: b,
                   value: endVertex(constEdges[0]).value,
                   edge: constEdges[0]) end
     where constEdges := from e: edgesFrom{Dataflow}(b)
                         with (b.commutative = false ? e.position = 1 : e.position <> -1)
                           and hasType{Const}(endVertex(e))
                         reportSet e end;
\end{lstlisting}

The query iterates over all \textsf{Binary} vertices that depend on at least
one constant.  The variable \textsf{constEdges} is bound to all
\textsf{Dataflow} argument edges leading to a \textsf{Const} with one
additional restriction: if the binary operation is not commutative, then the
immediate operation has to select the right operand (\textsf{position} = 1).
If the binary is commutative, then any operand may be used.

The query reports a set of records.  In each record, \textsf{binary} refers to
the \textsf{Binary} vertex, \textsf{value} refers to the value of the
\textsf{Const} operand's \textsf{value} attribute, and \textsf{edge} refers to
the the \textsf{Dataflow} edge that connects \textsf{binary} to the
\textsf{Const} operand.  The \textsf{Const} is deleted, and its value is pulled
into the immediate operation.

The interesting point is that the template graph specifies that the
\textsf{binary} has to be replaced with a new vertex whose type is specified by
a query instead of being fixed.  Here, the type name is calculated by
concatenating ``Target'' with the original binary's type name followed by
``I''.  Since both the type name and the reference query are optional, a type
name specified as query is denoted with a leading \textsf{\#}.

The \textsf{value} attribute is set to the \textsf{value} reported in each
match record.  The three dots (\textsf{...}) indicate that all other attributes
that are equally defined for the type of the replaced \textsf{binary} and the
new immediate binary have to be copied, e.g., the \textsf{associative} and
\textsf{commutative} values.  This saves one further operation call for
transforming \textsf{Cmp} vertices which are usual binaries with an additional
\textsf{relation} attribute.

The next operation replaces \textsf{MemoryNode} vertices
(\textsf{Load}/\textsf{Store}) that reference a \textsf{SymConst} to immediate
target load/store vertices.

\begin{lstlisting}[style=is]
MatchReplace (#'"Target" ++ typeName($.memory) ++ "I"' '$.memory' | symbol = '$.sym', ...)
  <== from m: V{MemoryNode},
           df: edgesFrom{Dataflow}(m),
           symConst: m --df-> & {SymConst}
      reportSet rec(memory: m, sym: symConst.symbol, df: df) end;
\end{lstlisting}

Again, the query returns a set of records.  The memory node \textsf{memory} is
replaced by a new immediate target memory node whose \textsf{symbol} attribute
is set to the value of the original \textsf{SymConst}'s \textsf{symbol} value.
The \textsf{Dataflow} edge \textsf{df} connecting the original memory node to
the \textsf{SymConst} is deleted.  Again, further attribute values are copied.

The two operations creating immediate target operations have created orphaned
\textsf{Const} and \textsf{SymConst} vertices, i.e., vertices of these types
that no other vertex depends on.  Thus, they can be delete.

\begin{lstlisting}[style=is]
Delete <== from c: V{Const, SymConst} with inDegree(c) = 0 reportSet c end;
\end{lstlisting}

Finally, the transformation replaces all other vertices whose type has a target
counterpart.

\begin{lstlisting}[style=is]
MatchReplace (#'"Target" ++ typeName($)' '$' | ...)
  <== V{^TargetNode, ^TargetMemoryNode, ^TargetMemoryNodeI,
        ^Block, ^Argument, ^Start, ^End, ^Phi, ^Return, ^Sync};
\end{lstlisting}

The query returns the set of all vertices that are not of type
\textsf{TargetNode}, \textsf{TargetMemoryNode}, \textsf{TargetMemoryNodeI},
\textsf{Block}, \textsf{Argument}, \textsf{Start}, \textsf{End}, \textsf{Phi},
\textsf{Return}, or \textsf{Sync}.  The former three exclude already
transformed vertices, and the rest are nodes of types that don't have a
different target type.

Again, in the template graph the type of the new vertex is specified with a
query.  Since the match query evaluates to a set of vertices in contrast to a
set of records, the variable \textsf{\$} holding the current match is just a
vertex which will be replaced while copying attribute values from the replaced
to the new vertex.

\section{Conclusion}
\label{sec:conclusion}

In this paper, the parts of the constant folding and and the complete
instruction selection transformation were explained.  With respect to the
evaluation criteria, both the \emph{Constant Folding} as well as the
\emph{Instruction Selection} solution are complete and correct, at least for
the provided test graphs.

With respect to performance, all given graphs could be transformed in times
between one and two seconds.  However, running the constant folding
transformation on the 100.000 nodes graph published shortly before the contest
takes about six minutes.  GReQL's set-based semantics of always calculating all
matching elements has considerable influence especially on the
\textsf{MatchReplace} operation.  This operation skips matches containing
elements of previous matches, because the previous application might have
changed them in a way that they would't match anymore.  Thus, in the worst case
any match except the first one has been calculated for nothing.

With respect to conciseness, about 190 lines of code including comments and
empty lines for three helpers and twelve operation calls is about as much as
one would expect for the constant folding transformation.  In contrast, the
instruction selection transformation with its 4 operation calls subsuming to 42
lines of code is very concise.

Concerning purity, the transformations are specified completely in GReTL.  But
the language has been extended with three in-place operations appropriate for
this task.  In this respect, both judging the solution as pure as well as
impure can be justified.  Nevertheless, it demonstrates GReTL's extensibility
which is a strength in its own respect.

\bibliographystyle{eptcs}
\bibliography{bibliography}

\begin{thebibliography}{1}
\providecommand{\bibitemdeclare}[2]{}
\providecommand{\urlprefix}{Available at }
\providecommand{\url}[1]{\texttt{#1}}
\providecommand{\href}[2]{\texttt{#2}}
\providecommand{\urlalt}[2]{\href{#1}{#2}}
\providecommand{\doi}[1]{doi:\urlalt{http://dx.doi.org/#1}{#1}}
\providecommand{\bibinfo}[2]{#2}

\bibitemdeclare{inproceedings}{compileroptimizationcase}
\bibitem{compileroptimizationcase}
\bibinfo{author}{Sebastian Buchwald} \& \bibinfo{author}{Edgar Jakumeit}
  (\bibinfo{year}{2011}): \emph{\bibinfo{title}{Compiler Optimization: A Case
  for the Transformation Tool Contest}}.
\newblock In \bibinfo{editor}{Pieter {Van Gorp}}, \bibinfo{editor}{Steffen
  Mazanek} \& \bibinfo{editor}{Louis Rose}, editors: {\sl
  \bibinfo{booktitle}{{TTC} 2011: Fifth Transformation Tool Contest, Z\"urich,
  Switzerland, June 29-30 2011}}, \bibinfo{publisher}{{EPTCS}}.

\bibitemdeclare{inproceedings}{tgapproach08}
\bibitem{tgapproach08}
\bibinfo{author}{J.~Ebert}, \bibinfo{author}{V.~Riediger} \&
  \bibinfo{author}{A.~Winter} (\bibinfo{year}{2008}):
  \emph{\bibinfo{title}{{G}raph {T}echnology in {R}everse {E}ngineering, {T}he
  {T}{G}raph {A}pproach}}.
\newblock In \bibinfo{editor}{R.~Gimnich}, \bibinfo{editor}{U.~Kaiser},
  \bibinfo{editor}{J.~Quante} \& \bibinfo{editor}{A.~Winter}, editors: {\sl
  \bibinfo{booktitle}{10th {W}orkshop {S}oftware {R}eengineering ({W}{S}{R}
  2008)}}, {\sl \bibinfo{series}{GI Lecture Notes in Informatics}}
  \bibinfo{volume}{126}, \bibinfo{publisher}{GI}, pp. \bibinfo{pages}{67--81}.

\bibitemdeclare{incollection}{FestschriftNagl2010}
\bibitem{FestschriftNagl2010}
\bibinfo{author}{J\"urgen Ebert} \& \bibinfo{author}{Daniel Bildhauer}
  (\bibinfo{year}{2010}): \emph{\bibinfo{title}{{Reverse Engineering Using
  Graph Queries}}}.
\newblock In: {\sl \bibinfo{booktitle}{Graph Transformations and Model Driven
  Engineering}}, \bibinfo{series}{LNCS 5765}, \bibinfo{publisher}{Springer},
  pp. \bibinfo{pages}{335--362}, \doi{10.1007/978-3-642-17322-6\_15}.

\bibitemdeclare{misc}{share-demo}
\bibitem{share-demo}
\bibinfo{author}{Tassilo Horn}: \emph{\bibinfo{title}{{SHARE} demo related to
  the paper {Solving the TTC 2011 Compiler Optimization Case with GReTL}}}.
\newblock
  \bibinfo{howpublished}{\url{http://is.ieis.tue.nl/staff/pvgorp/share/?page=C%
onfigureNewSession&vdi=Ubuntu_10.04_TTC11_gretl-cases.vdi}}.

\bibitemdeclare{inproceedings}{gretl-icmt2011}
\bibitem{gretl-icmt2011}
\bibinfo{author}{Tassilo Horn} \& \bibinfo{author}{J\"urgen Ebert}
  (\bibinfo{year}{2011}): \emph{\bibinfo{title}{The GReTL Transformation
  Language}}.
\newblock In \bibinfo{editor}{Jordi Cabot} \& \bibinfo{editor}{Eelco Visser},
  editors: {\sl \bibinfo{booktitle}{Theory and Practice of Model
  Transformations, Fourth International Conference, ICMT 2011, Zurich,
  Switzerland, June 27-28, 2011. Proceedings}}, {\sl \bibinfo{series}{Lecture
  Notes in Computer Science}} \bibinfo{volume}{6707},
  \bibinfo{publisher}{Springer}, pp. \bibinfo{pages}{183--197},
  \doi{10.1007/978-3-642-21732-6\_13}.

\end{thebibliography}

\clearpage
\appendix

\section{The Complete Solution}
\label{app:complete-solution}

In this appendix, the complete GReTL source code for the optional verifier, the
constant folding transformation, and the instruction selection transformation
is printed.

\subsection{The Verifier}
\label{app:verifier}

The verifier is not part of the challenge, but especially when developing a
transformation, a strict verification helps to spot errors.  GReTL does not
distinguish between libraries and transformations.  Any transformation like
\textsf{Verifier} (Listing~\ref{lst:verifier}) that does nothing except
declaring helper functions can be considered a library.  When executing such a
transformation, the helpers are made available to the calling transformation.

\begin{lstlisting}[language=gretl, float={h!t}, caption={The verifier as GReTL helper}, label={lst:verifier}]
transformation Verifier;

checkValidity() :=
  count(V{Start}) = 1 and count(V{StartBlock}) = 1             // (1)
  and count(V{End}) = 1 and count(V{EndBlock}) = 1             // (2)
  and (forall df: E{Dataflow},                                 // (3)
              hasType{Block}(endVertex(df))
       @ df.position = -1)
  and (forall n: V{Node,^Block}                                // (4)
       @ exists! df: edgesFrom{Dataflow}(n)
         @ df.position = -1 and hasType{Block}(endVertex(df)))
  and (forall c: V{Const}                                      // (5)
       @ c -->{Dataflow} theElement(V{StartBlock}))
  and (forall                                                  // (6)
         phi: V{Phi},
         block: phi-->{Dataflow @ thisEdge.position=-1} & {Block}
       @ outDegree{Dataflow}(phi) - 1 =                        // (6.1)
             outDegree{Controlflow}(block)
         and (forall                                           // (6.2)
                posNo: list(0..outDegree{Controlflow}(block) - 1)
              @ (exists! phiEdge: edgesFrom{Dataflow}(phi)
                 @ phiEdge.position = posNo)
                   and (exists! blockEdge:
                            edgesFrom{Controlflow}(block)
                        @ blockEdge.position = posNo)))
  and (forall block: V{Block,^EndBlock}                        // (7)
       @ inDegree(block) > 0)
  and (forall v: V @ degree(v) > 0);                           // (8)
\end{lstlisting}

In line 3, a GReQL helper named \textsf{checkValidity()} with no parameters is
defined.  It checks for eight constraints marked with numbered comments that
must hold for the current graph.

\begin{compactenum}[(1)]
\item There must be exactly one \textsf{Start} and one \textsf{StartBlock}.
\item There must be exactly one \textsf{End} and one \textsf{EndBlock}.
\item For all \textsf{Dataflow} edges pointing to a \textsf{Block}, the
  \textsf{position} attribute has to be set to -1.
\item For all \textsf{Node} vertices with the exception of \textsf{Block}
  vertices, there exists exactly one outgoing \textsf{Dataflow} edge with
  \textsf{position} set to -1 and leading to a \textsf{Block}, i.e., any node
  is contained in exactly one block.
\item All \textsf{Const} vertices are contained in the single
  \textsf{StartBlock}.
\item \textsf{Phi} vertices have the correct structure:
  \begin{compactenum}[(6.1)]
  \item \textsf{Phi} vertices have as many operand \textsf{Dataflow} edges as
    their containing \textsf{Block} has \textsf{Controlflow} predecessors.
  \item For any position number between 0 and the number of
    \textsf{Controlflow} predecessors of the \textsf{Phi}'s block minus one,
    there exists exactly one \textsf{Phi} argument edge with that position
    number and exactly one control flow predecessor with that position number.
  \end{compactenum}
\item There are no empty blocks.
\item There are no isolated vertices.
\end{compactenum}

Because the helper is a conjunction of predicates, its result is a boolean
value.  The transformations discussed in the next sections are asserting that
the helper returns \textsf{true} as first command (to ensure the given graph
is correct) and as last command (to ensure the transformation produced a valid
result).

\subsection{The Constant Folding Transformation}
\label{app:const-fold-transf}

In the following, the complete GReTL source code of the Constant Folding
transformation is shown.

\begin{lstlisting}[language=gretl]
transformation ConstantFolding;

// "import" the checkValidity() helper
ExecuteTransformation "transforms/verifier.gretl";

Assert <== checkValidity();

// Bind the single StartBlock to a variable
startBlock := theElement(V{StartBlock});

// Evaluates the given Binary with lval and rval.
evaluateBinary() := using bin, lval, rval:
    hasType{Add}(bin)  ? lval + rval
  : hasType{Sub}(bin)  ? lval - rval
  : hasType{Mul}(bin)  ? lval * rval
  : hasType{Div}(bin)  ? (let d := lval / rval in d < 0 ? ceil(d) : floor(d))
  : hasType{Mod}(bin)  ? lval % rval
  : hasType{Shl}(bin)  ? bitOp("SHIFT_LEFT", lval, rval)
  : hasType{Shr}(bin)  ? bitOp("UNSIGNED_SHIFT_RIGHT", lval, rval)
  : hasType{Shrs}(bin) ? bitOp("SHIFT_RIGHT", lval, rval)
  : hasType{And}(bin)  ? bitOp("AND", lval, rval)
  : hasType{Or}(bin)   ? bitOp("OR", lval, rval)
  : hasType{Eor}(bin)  ? bitOp("XOR", lval, rval)
  : hasType{Cmp}(bin)  ?
        (bin.relation = "GREATER"        ? (lval > rval  ? 1 : 0)
       : bin.relation = "GREATER_EQUALS" ? (lval >= rval ? 1 : 0)
       : bin.relation = "LESS"           ? (lval < rval  ? 1 : 0)
       : bin.relation = "EQUAL"          ? (lval = rval  ? 1 : 0)
       : bin.relation = "NOT_EQUAL"      ? (lval <> rval ? 1 : 0)
       : bin.relation = "LESS_EQUAL"     ? (lval <= rval ? 1 : 0)
       : bin.relation = "TRUE"           ? 1
       : bin.relation = "FALSE"          ? 0
       : error("Don't know how to handle " ++ bin.relation))
  : error("Don't know how to handle " ++ bin);

// Sorts the given dataflows by the value of their position attribute.
sortByPosition() := using dataflows:
  let posMap := from i: list(0..count(dataflows) - 1)
                reportMap dataflows[i].position -> dataflows[i] end,
      sortedKeys := sort(keySet(posMap))
  in from i: sortedKeys
     reportSet posMap[i] end;

// A helper that returns all non-containment dataflows of a node.  That means,
// dataflows where position is not -1.
nonContainmentDfs() := using node:
  from e: edgesFrom{Dataflow}(node)
  with e.position <> -1
  reportSet e end;

// Perform the contained transformations as long as one was applicable
Iteratively
  // Replace associative, commutative Binaries of form ((x + y) + 1) with ((x +
  // 1) + y), that is, pull up constants.
  Iteratively MatchReplace b1('$.b1'), b2('$.b2'),
                           b1 -->{'$.df1'} b2 -->{'$.df3'} ('$.c'),
                           b2 -->{'$.df4'} ('$.y'),
                           b1 -->{'$.df2'} ('$.x')
    <== from b1: V{Add, Mul},
             df1, df2: nonContainmentDfs(b1),
             b2: b1 --df1-> & {Add, Mul @ type(b1) = type(thisVertex)},
             df3, df4: nonContainmentDfs(b2)
        with  df1 <> df2 and df3 <> df4
          // Used only by b1 + the containment Dataflow
          and inDegree{Dataflow}(b2) = 2
          and hasType{Const}(endVertex(df2))
          and not(hasType{Const}(x))
        reportSet rec(b1: b1, b2: b2, x: x, c: theElement(b1 --df2->),
                      y: y, df1: df1, df2: df2, df3: df3, df4: df4) end
        where x := theElement(b2 --df3->),
              y := theElement(b2 --df4->);;

  // Replace Binary with 2 Consts with a single Const of the result.  Do
  // iteration here, cause that operation is likely to produce another match
  // immediately.
  Iteratively MatchReplace ('startBlock') <--{Dataflow | position = '-1'}
                           (Const '$.op' | value = '$.value')
    <== from binOp: V{Binary}
        with hasType{Const}(lconst) and hasType{Const}(rconst)
        reportSet rec(op: binOp,
                      value: evaluateBinary(binOp, lconst.value, rconst.value),
                      edges: edgesFrom(binOp)) end
        where argEdges := sortByPosition(nonContainmentDfs(binOp)),
              lconst := endVertex(argEdges[0]),
              rconst := endVertex(argEdges[1]);;

  // Replace unary ops (i.e., Not) with the value
  Iteratively MatchReplace ('startBlock') <--{Dataflow | position = '-1'}
                           (Const '$.op' | value = '$.value')
    <== from uniOp: V{Not}
        with hasType{Const}(endVertex(argEdge))
        reportSet rec(op: uniOp,
                      value: bitOp("NOT", endVertex(argEdge).value),
                      edges: edgesFrom(uniOp)) end
        where argEdge := theElement(nonContainmentDfs(uniOp));;

  // Remove unused Consts
  Delete <== from c: V{Const} with inDegree(c) = 0 reportSet c end;

  // Merge duplicate Consts into one for each value.
  MergeVertices <== from const: V{Const}
                    reportMap const -> from d: V{Const}
                                       with d.value = const.value
                                         and d <> const
                                       reportSet d end
                    end;

  // Remove duplicate edges from StartBlock to Consts.
  Delete <== from c: V{Const}
             reportSet difference(dfs, set(dfs[0])) end
             where dfs := edgesFrom{Dataflow}(c);

  // Replace Cond nodes depending only on a Const with an unconditional Jmp.
  MatchReplace ('$.block') <--{'$.df'} (Jmp) <--{Controlflow} ('$.target')
    <== from cond: V{Cond}
        with not isEmpty(cond -->{Dataflow} & {Const})
        reportSet rec(cond: cond,           // the Cond
                      df: blockTup[0],      // the Dataflow to the containing Block
                      block: blockTup[1],   // the containing Block
                      cf: targetTup[0],     // the True/False edge to the target
                      target: targetTup[1], // the target of the Cond
                      edges: edgesConnected(cond)) end
        where const := theElement(cond -->{Dataflow} & {Const}),
              blockTup := theElement(from e: edgesFrom{Dataflow}(cond)
                                     with e.position = -1
                                     reportSet e, endVertex(e) end),
              targetTup := theElement(from e: edgesTo{Controlflow}(cond)
                                      with const.value = 0 ? hasType{False}(e)
                                                           : hasType{True}(e)
                                      reportSet e, startVertex(e) end);

  // Now delete any node that has no outgoing Dataflow recursively
  Iteratively Delete <== from node: V{^StartBlock}
                         with outDegree(node) = 0
                         reportSet node end;;

  // Remove the superfluous operand edges of Phis
  Delete <== from phi: V{Phi}, opEdge: edgesFrom{Dataflow}(phi)
             with opEdge.position <> -1
               and not(contains(predBlockCfPositions, opEdge.position))
             reportSet opEdge end
             where predBlock := theElement(phi -->{Dataflow @ thisEdge.position = -1}),
                   predBlockCfPositions := from df: edgesFrom{Controlflow}(predBlock)
                                           reportSet df.position end;

  // Renumber Phi operand edges and the corresponding block predecessor edges.
  MatchReplace ('$.phiOperand') <--{'$.phiOpEdge' | position = '$.position'} ('$.phi'),
               ('$.blockPred') <--{'$.blockCfEdge' | position = '$.position'} ('$.block')
    <== from phi: V{Phi}, i: list(0..count(blockEdges) - 1)
        reportSet rec(phiOperand: endVertex(phiOpEdges[i]),
                      phiOpEdge: phiOpEdges[i],
                      position: i,
                      phi: phi,
                      block: block,
                      blockCfEdge: blockEdges[i],
                      blockPred: endVertex(blockEdges[i])) end
        where phiOpEdges := sortByPosition(nonContainmentDfs(phi)),
              block := theElement(phi -->{Dataflow @ thisEdge.position = -1}),
              blockEdges := sortByPosition(edgesFrom{Controlflow}(block));

  // Remove Phis with only one argument
  MatchReplace ('$.arg') <--{'$.depEdge'} ('$.dep')
    <== from phi: V{Phi}
        with count(argEdges) = 1
        reportSet rec(arg: theElement(phi --argEdge->),
                      depEdge: userEdge,
                      dep: theElement(phi <-userEdge--),
                      phi: phi,
                      edges: edgesConnected(phi)) end
        where argEdges := nonContainmentDfs(phi),
              argEdge  := argEdges[0],
              userEdge := theElement(edgesTo{Dataflow}(phi));

  // Remove Blocks that contain only a single Jmp
  MatchReplace ('$.pred') <--{'$.predEdge'} ('$.target')
    <== from block: V{Block}, jmp: V{Jmp}
        with block <--{Dataflow @ thisEdge.position = -1} jmp
          and inDegree(block) = 1
        reportSet rec(pred: theElement(block -->{Controlflow}),
                      predEdge: theElement(edgesFrom{Controlflow}(block)),
                      target: theElement(jmp <--{Controlflow}),
                      edges1: edgesConnected(block),
                      edges2: edgesConnected(jmp),
                      block: block) end;

; // End outer Iteratively

// Finally, check if still everything's correct.
Assert <== checkValidity();
\end{lstlisting}

\subsection{The Instruction Selection Transformation}
\label{app:instr-select-transf}

In the following, the complete GReTL source code of the Instruction Selection
transformation is shown.

\begin{lstlisting}[language=gretl]
transformation InstructionSelection;

// "import" and execute the checkValidity() helper
ExecuteTransformation "transforms/verifier.gretl";
Assert <== checkValidity();

// Transform all Binaries with at least one Const to an immediate TargetBinary
// operation.
MatchReplace (#'"Target" ++ typeName($.binary) ++ "I"' '$.binary'| value = '$.value', ...)
  <== from b: V{Binary}
      with count(constEdges) > 0
      reportSet rec(binary: b,
                    value: endVertex(constEdges[0]).value,
                    edge: constEdges[0]) end
      where constEdges := from e: edgesFrom{Dataflow}(b)
                          with (b.commutative = false ?
                                 // For non-commutative binary ops, we have to chose the
                                 // dataflow on position 1, i.e., the second argument
                                 e.position = 1
                                 // ...else anything except -1 is ok.
                                : e.position <> -1)
                            and hasType{Const}(endVertex(e))
                          reportSet e end;

// Transform Load/Store with SymConsts to TargetLoadI/TargetStoreI.
MatchReplace (#'"Target" ++ typeName($.memory) ++ "I"' '$.memory' | symbol = '$.sym', ...)
  <== from m: V{MemoryNode},
           df: edgesFrom{Dataflow}(m),
           symConst: m --df-> & {SymConst}
      reportSet rec(memory: m, sym: symConst.symbol, df: df) end;

// Delete unused Const/SymConst vertices.
Delete <== from c: V{Const, SymConst} with inDegree(c) = 0 reportSet c end;

// Transform the rest of the elements.  TargetMemoryNode and TargetMemoryNodeI
// have to be excluded explicitly, because they are not derived from TargetNode.
MatchReplace (#'"Target" ++ typeName($)' '$' | ...)
  <== V{^TargetNode, ^TargetMemoryNode, ^TargetMemoryNodeI,
        ^Block, ^Argument, ^Start, ^End, ^Phi, ^Return, ^Sync};

// Finally, check if still everything's correct.
Assert <== checkValidity();
\end{lstlisting}

\end{document}